\documentclass[a4paper,11pt]{article}
\usepackage{pos}
\usepackage[textwidth=25mm,textsize=small]{todonotes}
\usepackage{tikz}

\newcommand{\package}[1]{\texttt{#1}}
\newcommand{\code}[1]{\texttt{#1}}
\newcommand{\dir}[1]{\texttt{#1}}

\newcommand{\gls}[1]{#1}

\newcommand{\drbar}{\ensuremath{\overline{\text{DR}}}}
\newcommand{\msbar}{\ensuremath{\overline{\text{MS}}}}

\newcommand{\observable}{\dir{O$_{i}$}}

\newcommand{\wl}{\code{Wolfram Language}}
\newcommand{\cpp}{\code{C++}}
\title{NPointFunctions: a calculator of amplitudes and observables in FlexibleSUSY}

\usepackage{orcidlink}
\author*{Uladzimir Khasianevich\,\orcidlink{0000-0003-0255-0674}}
\renewcommand{\printHeadAuthors}{Uladzimir Khasianevich}
\author{Wojciech Kotlarski\,\orcidlink{0000-0002-1191-6343}}
\author{Dominik St\"ockinger}

\affiliation{%
	Institut f\"ur Kern- und Teilchenphysik, TU Dresden,\\ 
	Zellescher Weg 19, 01069 Dresden, Germany%
}

\emailAdd{uladzimir.khasianevich@tu-dresden.de}
\emailAdd{wojciech.kotlarski@tu-dresden.de}
\emailAdd{dominik.stoeckinger@tu-dresden.de}

\abstract{%
We present \package{NPointFunctions}, a package to incorporate desired amplitudes and observables for an arbitrary BSM model in \package{FlexibleSUSY}.
The package relies on the SARAH-generated output used with \package {FeynArts}/\package{FormCalc}, embedded in an appropriate way, thus
allowing calculations up to one-loop level.
The resulting tool is an extension to \package{FlexibleSUSY}, a generator of spectrum generator programs.
\package{NPointFunctions} was designed to be customizable, modular and extensible with additional process- or amplitude- dependent contributions.
We explain, how these goals are achieved, and discuss
modifications to the default \package{FlexibleSUSY} workflow.
}

\FullConference{%
  Computational Tools for High Energy Physics and Cosmology (CompTools2021)\\
  22-26 November 2021\\
  Institut de Physique des 2 Infinis (IP2I), Lyon, France
}

\begin{document}
\maketitle
\section{Introduction}
In the desire to study the parameter space of some \gls{BSM} theories, one
often uses software packages.
However, they usually exist for a very narrow range of models, mostly very close to \gls{SM}, \gls{MSSM}, \gls{2HDM}, and some higher dimensional operator
additions.
This is only a slice of all interesting possibilities, so that \package{FlexibleSUSY}~\cite{Athron:2014yba, Athron:2017fvs} was created
to be used for an arbitrary \gls{SUSY} or non-\gls{SUSY} model.%
\footnote{The only code with a similar capability is \package{SARAH}/\package{SPheno}/\package{FlavorKit}~\cite{Staub:2013tta,Porod:2003um,Porod:2011nf,Porod:2014xia}.}
It is a program written mostly in \wl{} and \cpp{}, which generates a fast and precise \cpp{} spectrum generator for any model specified by the user.
The output spectrum-program applies user-defined boundary conditions at up to three different energy scales for the model (thus utilizing \gls{RGE} running between them) and creates a set of mixing matrices, pole masses, and some auxiliary quantities, which might be specified in the corresponding \package{FlexibleSUSY} model file.
Recent versions of \package{FlexibleSUSY} introduced the computation of several important observables which are suitable for phenomenological studies and comparisons against experimental data; we highlight in particular the extension \package{FlexibleDecay}~\cite{Athron:2021kve}.

To simplify and modularize the addition of new observables to \package{FlexibleSUSY}, an extension named \package{NPointFunctions} was created.
This package is an engine for the automated calculation of amplitudes and quantities that rely on them for any model.
It incorporates a well-defined approach of widely used  packages  \package{FeynArts}~\cite{Hahn:2000kx} and \package{FormCalc}~\cite{Hahn:1998yk} up to some technical details to be mentioned later.
It was designed and structured to be:
\begin{description}
\item[customizable] as the user can access and modify the calculation of observables with the help of simple and advanced settings; 
\item[extensible] as one can straightforwardly add new or change existing observables;
\item[modular] as the file structure of the \wl{} and \cpp{} code transparently reflects performed tasks.
\end{description} 

In the following sections, we briefly describe the changes of the main \package{FlexibleSUSY} workflow and show how design goals mentioned above are implemented.
\section{Ways to calculate and customize \code{NPointFunctions} observables}
In this section, we describe what the user should change to calculate an observable and how the calculation of the observable can be modified.

To create a \cpp{} spectrum generator, \package{FlexibleSUSY} usually goes through the so-called meta-phase.
It executes \wl{} routines of the \dir{meta} directory, which add model-specific information, produced by \package{SARAH}, to template files in \cpp{} language of the	 \dir{templates} directory.
The resulting set of filled \cpp{} files is combined into a single program \textemdash{} a spectrum generator. 

The decision about observables to calculate is done during the meta-phase.
To add them the user needs to change the settings file \dir{FlexibleSUSY.m.in} of the corresponding model.
Namely, one adds desired observables into \code{ExtraSLHAOutputBlocks}. 
Once added, the observables modify the standard meta-phase steps, in particular, by launching \package{NPointFunctions}. 

Different observables are defined by a unique set of options, which usually contain: 
generations of external particles,
the loop level,
the combination of diagrams to include,
the type of output, etc.
Up to one-loop level amplitudes can be calculated due to \code{FormCalc}, however in the case of higher loops, one can include additional external packages and/or use some hard-coded expressions, like in other \code{FlexibleSUSY} extensions.
By diagrams we mean something very general, like penguins and boxes with all fields or penguins without tree-level scalar fields (which might be useful for Higgs particle propagation).
The supported output formats are \code{SLHA}~\cite{Skands:2003cj,Allanach:2008qq},
\code{FLHA}~\cite{Mahmoudi:2010iz}, and \code{WCxf}~\cite{Aebischer:2017ugx}.

If more customization for an observable is required, then one can use advanced ways, designed to be straightforward for the user, to achieve it.
Some general \package{NPointFunctions} settings, specific for the observable, can be changed in the appropriate \dir{\observable/class.m} file (see figure~\ref{fig:structure}).
They include \msbar{}/\drbar{} renormalization scheme, external momenta treatment, usage of on-shell external particles, which mode of \package{NPointFunctions} to use, etc.
Currently, there are two modes to calculate an amplitude: using general settings only and using additional options in a \dir{\observable/settings.m} file.
The former is used so far for cross-checks via the unit-test framework for several self energies in the \gls{SM}/\gls{MSSM} and Z-boson penguins in the \gls{MRSSM}~\cite{Kribs:2007ac}.
The latter allows to customize plenty of different options for the calculation
of the observable.

As it was mentioned previously, we rely on \code{FeynArts}/\code{FormCalc} to create amplitudes.
However, due to the usage of \cpp{} template metaprogramming and the specifics of required amplitudes, we use the packages in a non-standard way.
For example, the particle content of the model, vertices, and masses are stored in appropriate namespaces. 
This allows us to use generic expressions of amplitudes only and make a \cpp{} compiler substitute all classes with appropriate particle generations. 
Thus the amplitudes are modified appropriately before the usage of a \code{FormCalc} routine because it calculates amplitudes at the level of particles, which is not required by \code{FlexibleSUSY}.
It also requires us to obtain all color factors for amplitudes by other means and we rely here on \package{ColorMath}~\cite{Sjodahl:2012nk}.

After these introductory remarks, we explain the \dir{\observable/settings.m} file, which can be created for every observable \observable{} and contains low-level instructions for \package{FeynArts}/\package{FormCalc}.
We highlight only the most important settings: \code{topologies[LOOPS]},
\code{diagrams[LOOP, TYPE]}, \code{amplitudes[LOOP, TYPE]}, \code{chains[LOOPS]}, \code{sum[LOOPS]}.
The first three allow to exclude unwanted contributions based on: the adjacency matrix of a diagram, generic fields in a diagram, and class insertions.
The fourth specifies the fermionic operators to neglect.
The last one defines which particle generations to skip at the \cpp{} level. 

The structure of these settings is naturally defined by how topologies, diagrams, and amplitudes are stored internally and will be explicitly explained in the forthcoming manual.
A tree-shaped container allows us to use its descending structure to simply remove undesired combinations even for some \emph{amplitude} substitutions.
Also, settings responsible for \emph{changes} of amplitudes usually depend on the topology where they should be applied, which is naturally reflected by the tree.
Let us now shed light on the details of the mentioned settings.

The entry point for the calculation is always defined by \code{topologies[LOOP]}, which defines the required graphs and which is a set of connections between user-defined symbols and topology names.
The latter uniquely defines topologies by corresponding connections to adjacency matrices.
An integer variable \code{LOOP} defines the loop level of the observable \observable{}, where these settings can be applied.

Generic fields are inserted into topologies by \package{FeynArts}.
One uses \code{diagrams[LOOP, TYPE]} to define generic fields which should not be inserted. 
The variable \code{TYPE} defines whether the option should be applied if some setting is present in \code{FlexibleSUSY.m.in} file or not.
For example, we have defined setting names \code{Vectors} and \code{Scalars}.
The first one is aimed to be used for diagrams with vector bosons only and the second only with scalars.
Mentioning \code{Vectors} in the corresponding observable in \code{FlexibleSUSY.m.in} assumes that there are no diagrams with scalars in the output and vice versa.
If we use the names separately, then it is useful to make \code{Vectors} remove all
vectors if it is unused and similarly for the \code{Scalar} name.
Then, it allows us to combine both settings and keep both vectors and scalars without additional chaos of checks for a different combination of options.

Having expressions of diagrams allows us to calculate amplitudes with \package{FormCalc}, which makes generic field substitutions for class levels.
One specifies in \code{amplitudes[LOOP, TYPE]} the contributions to exclude, based on class insertions.
For example, amplitudes with massless particles usually require special treatment and this option allows to remove them from the generated expressions in order to handle them in an appropriate way.

One of the goals for \package{NPointFunctions} is to provide a simple way to extract Wilson coefficients.
If one considers observables with external fermions, then there might exist a need to change or fine-tune fermionic chains in order to achieve the structures
of the desired subset of operators.
It is done by \code{chains[LOOP]}, where chains to be dropped are specified. 

In some cases, one can't sum over all generations for some generic field in the amplitude.
For example, in \gls{cLFV} observables, one usually has self-energy-like diagrams within the so-called penguin contribution.
The fermion in a tree-like propagator should differ from the external one. 
This behavior can be defined in \code{sum[LOOP]}.  

There are other settings of minor importance.
More settings, as well as observables, can be added to \package{NPointFunctions} by the user.
It is rather straightforward thanks to the modular structure described in the next section.
\section{The modular structure of \code{NPointFunctions}}
\begin{figure}[t!]
\centering
\usetikzlibrary{fit,backgrounds}
\begin{tikzpicture}[grow'=right,
	edge from parent path=
		{(\tikzparentnode.east) |- (\tikzchildnode.west)},
	level 1/.style={level distance=0cm, sibling distance=0.8cm},
	level 2/.style={level distance=3.5cm, sibling distance=.5cm},
	level 3/.style={level distance=3.8cm},
	level 4/.style={level distance=3cm},
	back/.style={inner sep=1mm, rounded corners=3mm,
		draw=gray!50, fill=gray!5},
]
\node {}
	child[sibling distance=0.65cm] { node (O) {\dir{Observables}}
		child { node (o) {\dir{observable}} } edge from parent [draw=none] }
	child { node (W) {\dir{WriteOut}} 
		child { node (w) {\dir{write}} } edge from parent [draw=none] }
	child { node (F) {\dir{FlexibleSUSY}} 	
		child { node (m) {\dir{class}}
			child { node (N) {\dir{NPointFunctions}} } } 
		child { node (s) {\dir{[settings]}}
			child { node (i) {\dir{internal}}
				child { node (s*) {\dir{settings$^*$}} }
				child { node (r) {\dir{rules}} } } edge from parent [draw=none] } edge from parent [draw=none] }
	child { node (M) {\dir{FSMathLink}} 
		child { node (l) {\dir{[librarylink]}} } edge from parent [draw=none] }
;
\begin{pgfonlayer}{background}
\node [back, fit=(O) (W) (F) (M),
	label={[black!60]above:\package{FlexibleSUSY}},
	label={[black!60]below:\dir{configure}}] {}
;
\node [back, fit=(o) (w) (m) (s) (l),
	label={[black!60]above:Observable \dir{\observable}},
	label={[black!60]below:\dir{FlexibleSUSY.m.in}}] {}
;
\node [back, fit=(N) (i) (s*) (r),
	label={[black!60]above:\package{NPointFunctions}},
	label={[black!60]below:\dir{\observable/class.m}}
	] {}
;
\end{pgfonlayer}
\end{tikzpicture}
\caption{
	A subset of \dir{.m} files and their dependencies, which are evaluated during the \package{FlexibleSUSY} meta phase.
	They are relevant for observable \dir{O$_i$}, generated with the help of the \package{NPointFunctions} extension.
	The file names (\dir{configure}, \dir{FlexibleSUSY.m.in} and \dir{\observable/class.m}) under highlighted blocks define whether corresponding block is evaluated or not.
}
\label{fig:structure}
\end{figure}
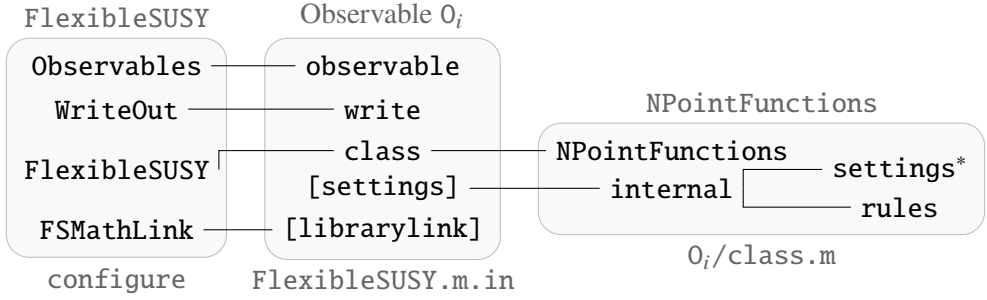

In the previous section, we have briefly discussed the default meta-phase of \package{FlexibleSUSY} and mentioned that the user can include the observable \observable{} and modify its settings.
Now we highlight the most important steps of the default meta-phase, as indicated by the \package{FlexibleSUSY} block in figure~\ref{fig:structure}, and describe the changes introduced by \package{NPointFunctions}.
These seemingly redundant details are required to \emph{add} a new observable \observable{}: one needs to create a separate file for each of the meta-phase steps in \dir{meta/NPointFunctions/\observable} directory: \dir{observable.m}, \dir{write.m}, \dir{class.m} and two \cpp{} template files in \code{templates/npointfunctions} directory.
If some advanced settings for \package{FeynArts/FormCalc} are required, then \dir{settings.m} can be created.
The content of the last file was described in the previous section.
More knowledge is required to write the other ones.

The default meta-phase of \package{FlexibleSUSY} performs several steps to convert \code{SARAH} output into a \cpp{} spectrum generator.
The only important ones are shown in the \package{FlexibleSUSY} block:
\begin{description}
\item[\dir{Observables.m}] connects the name of observable given in the \dir{FlexibleSUSY.m.in} file with the \cpp{} definitions: name, textual description, type, and a function name to calculate the observable are listed there.
This information allows \package{FlexibleSUSY} to fill template files correctly.
To add a new observable a separate file \dir{observable.m} should be created with the data specified above.

\item[\dir{WriteOut.m}] specifies, how to output the result of calculation in a way that can be read and passed to other programs via the \code{SLHA} and \code{FLHA} formats.
In particular, the new observable file \dir{write.m} defines the exact structure of the Wilson coefficients which are calculated by the observable.

\item[\dir{FlexibleSUSY.m}] fills \cpp{} template files with model-specific information. 
Each file is modified by a specific "write class" function.
For a new observable one defines a similar one in \dir{class.m} file.
If there is a need to use \package{FeynArts}/\package{FormCalc}, then
it should be specified inside this file.
It allows adding new packages, capable of doing calculations with more loops, or replacing used ones in the future.
Also, to fill template files for an observable, one needs to create them in \code{templates/npointfunctions} directory. 
The only requirement for them is to contain the definition of the calculate function defined in \dir{observable.m}.

\item[\dir{FSMathLink.m}] defines, how to call observable via the \wl{} interface. 
One can create an optional file \dir{librarylink.m} and specify there the desired interface.
\end{description}

If the \observable{} is mentioned appropriately in \code{FlexibleSUSY.m.in}, then all files of the Observable \observable{} block on figure~\ref{fig:structure} are loaded during the \package{FlexibleSUSY} meta-phase automatically.

One can define new settings for \package{NPointFunctions}.
The modular structure of the package, shown in the \package{NPointFunctions} block in figure~\ref{fig:structure}, makes it transparent:
\begin{description}
\item[\code{NPointFunctions.m}] is an entry point for the extension that specifies in which mode \package{FeynArts} and \package{FormCalc} should be used and translates amplitudes in \package{SARAH} conventions into \cpp{} code.

\item[\code{internal.m}] operates with \package{FeynArts} and \package{FormCalc} in their specific conventions, loading all other \code{settings*.m} parser files as well. 

\item[\code{settings*.m}] is a set of files under different names.
Each file describes the appropriate operations for a corresponding setting.
It's simpler to keep a parser of every specific setting in a separate file because then the modifications are orthogonal.

\item[\code{rules.m}] contains a set of conversion rules from \code{FeynArts}/\package{FormCalc} conventions to \package{SARAH} ones.
\end{description}
\section{Conclusions}
We have provided a brief overview of a new extension to the \package{FlexibleSUSY} program, called \package{NPointFunctions}.
It allows adding the automatic computation of observables to spectrum generators generated by \package{FlexibleSUSY}.
The observables can be computed up to the one-loop level. 
Important design goals are customizability, extensibility, and modularity of the code. 
We have described the basic structure of the code, reflecting these goals, and we have illustrated the possibilities of code modification by the user via settings files and by the addition of new observables and new options to the \package{FeynArts}/\package{FormCalc} packages used during their evaluation.
A more detailed manual with specific instructions and usage examples will be the
subject of a forthcoming publication. 
\acknowledgments
We thank the \package{FlexibleSUSY} authors for helpful discussions.
In particular, U.Kh. thanks Jobst Ziebell~\cite{Ziebell:2018th} and Kien Dang Tran~\cite{DangTran:2019th} for creation of the initial version of the \package{NPointFunctions}.
U.Kh. was supported by the Deutscher Akademischer Austauschdienst (DAAD) under Research Grants --- Doctoral Programmes in Germany, 2019/20 (57440921) and by
the German Research Foundation (DFG) under grant number STO 876/7-1.
The work was supported by the DFG under grants number STO 876/4-1 and STO 876/2-2.

\bibliographystyle{jhep.bst}
\bibliography{bibliography}
\end{document}